\documentclass[journal]{IEEEtran}
\usepackage{cite}
\usepackage{amsmath,amssymb,amsfonts,amsthm}
\usepackage{algorithmic}
\usepackage{graphicx}
\usepackage{textcomp}
\usepackage{caption}
\usepackage{subfigure}
\usepackage{booktabs}
\usepackage[colorlinks=true, allcolors=blue]{hyperref}
\usepackage[linesnumbered,ruled,vlined]{algorithm2e}
\usepackage{wrapfig}
\usepackage{cite}
\usepackage{xr}
\usepackage{hyperref}
\usepackage[normalem]{ulem}

\makeatletter
\newcommand*{\addFileDependency}[1]{
  \typeout{(#1)}
  \@addtofilelist{#1}
  \IfFileExists{#1}{}{\typeout{No file #1.}}
}
\makeatother

\usepackage{adjustbox}
\usepackage{setspace}
\usepackage{hyperref}
\usepackage{multirow}
\usepackage[switch,columnwise]{lineno}
\usepackage{chapterbib}

\newcommand{\etal}{\textit{et al}.}

\SetKwInput{KwInput}{Input}                
\SetKwInput{KwOutput}{Output}              

\begin{document}

\title{Assessing the Impact of Deep Neural Network-based Image Denoising on Binary Signal Detection Tasks}

\author{Kaiyan Li, Weimin Zhou, Hua Li, and Mark A. Anastasio 

\thanks{
Kaiyan Li is with the Department of Bioengineering, University
	of Illinois at Urbana-Champaign, Urbana, IL, 61801 USA (Email:
	kaiyanl2@illinois.edu).
Weimin Zhou was with the Department of Electrical and Systems
	Engineering, Washington University in St. Louis, St. Louis, MO, 63130
	USA. He is now with the Department of Psychological and Brain Sciences, University of California, Santa Barbara, Santa Barbara, CA 93106 USA (Email: weiminzhou@ucsb.edu).
Hua Li is with the Department of Bioengineering, University of Illinois
	at Urbana-Champaign, Urbana, IL 61801 USA, and with the Carle
	Cancer Center, Carle Foundation Hospital, Urbana, IL 61801 USA (Email:
	huali19@illinois.edu).
Mark A. Anastasio is with the Department of Bioengineering, University
	of Illinois at Urbana-Champaign, Urbana, IL, 61801 USA (Email:
	maa@illinois.edu).
This work was supported in part by NIH awards R01EB020604, R01EB023045, R01NS102213, R01CA233873, and R21CA223799. (Corresponding authors: Mark A. Anastasio \& Hua Li).
}
\vspace{-0.6cm}
}
\maketitle

\begin{abstract}
A variety of deep neural network (DNN)-based image denoising methods have been proposed for use with medical images.
Traditional measures of image quality (IQ) have been employed to optimize and evaluate these methods.
However, the objective evaluation of IQ for the DNN-based denoising methods remains largely lacking.
In this work, we evaluate the performance of DNN-based denoising methods
by use of task-based IQ measures. 
Specifically, binary signal detection tasks under signal-known-exactly (SKE) with background-known-statistically (BKS) conditions are considered.
The performance of the ideal observer (IO) and common linear numerical observers are quantified 
and detection efficiencies are computed 
to assess the impact of the denoising operation on task performance.
The numerical results indicate that, in the cases considered, the application of a denoising network
can result in a loss of task-relevant information in the image. 
The impact of the depth of the denoising networks on task performance is also assessed.
The presented results highlight the need for the objective evaluation of IQ for DNN-based denoising technologies
and may suggest future avenues for improving their effectiveness in medical imaging applications.
\end{abstract}


\section{Introduction}
\label{sec:introduction}  

Image denoising is a classical image processing operation that is commonly employed  in medical imaging applications~\cite{manduca2009projection,li2014adaptive,lin2001improving,le2013denoising,gong2018pet,chen2017low,zhang2017beyond}.  
Recently, denoising methods based  on deep neural networks (DNNs) have been proposed 
and widely investigated~\cite{yang2018low,gondara2016medical,gong2018pet, chen2017low,kang2017deep,chen2017low2,liu2019deep,you2019denoising,liu2020connecting,tian2018deep,you2018structurally,you2019low}.
These methods are typically trained by minimizing loss functions that quantify a distance between the denoised image and the defined target image (e.g., a noise-free or low noise image)
and have demonstrated high performance 
in terms of traditional image quality metrics such as root mean square error (RMSE), 
structural similarity index metric (SSIM)~\cite{wang2004image} or peak signal-to-noise ratio (PSNR). 

In medical imaging, images are often acquired for specific purposes and the use of objective measures of image quality (IQ) has been widely advocated for assessing imaging systems and image processing algorithms \cite{barrett2013foundations,metz1995toward,vennart1997icru,wagner1985unified,he2013model,li2021supervised}.
Despite this, the objective evaluation of  modern DNN-based medical image denoising methods remains largely lacking \cite{yu2020ai}. 
Although DNN-based denoising methods, by conventional design, can improve traditional IQ measures, it is well-known that such measures may not always correlate with objective task-based IQ measures~\cite{barrett1993model,christianson2015improved,myers1985effect,badal2019virtual,li2021task}. For example, Yu \emph{et al.} \cite{yu2020ai} conducted a  study in which  a DNN-based denoising method was observed to reduce RMSE  compared to an alternative method, but signal detectability was unimproved \cite{yu2020ai}. 

Even more concerning is the fact that image denoising methods can compromise 
the visibility of important structural details in the denoised images even though traditional measurement metrics (such as RMSE or SSIM) are improved~\cite{yang2018low,li2020mri,li2014adaptive}. 
While DNN-based denoising operations may succeed at lowering noise levels, 
the extent to which they perturb the second- and higher-order statistical properties of an image that are relevant to signal detection is not understood.  Finally, according to data processing inequality~\cite{beaudry2012intuitive}, 
the performance of an ideal observer cannot be increased via image processing operations such as denoising. However, conditions under which DNN-based denoising methods can improve the performance of sub-optimal observers on detection tasks remains relatively unexplored.

The purpose of this study is to assess modern DNN-based denoising methods by use of objective IQ measures, in a preliminary attempt to address the issues described above. Three canonical DNN-based denoising methods are identified for analysis.
The convolutional neural network (CNN)-based observer, the Hotelling observer, the Regularized Hotelling observer, an anthromorphic channelized Hotelling observer, and a non-prewhitening matched filter are implemented as NOs.
The performances of these NOs acting on the original noisy images and the corresponding denoised images are quantified via receiver operating characteristic (ROC) analysis, and
signal detection efficiencies are computed to assess the impact of the denoising operations on  NO performance. The impact of the network depth of a DNN-based denoising method on NO performance is assessed to understand if the deep learning mantra ``deeper is better" necessarily holds true for signal detection performance.
A noise propagation analysis is also performed, to gain insights into how DNNs modify the covariance structure of image data as they are propagated through layers of a linear convolutional network. 
Finally, the depth of the CNN-based observer is varied to demonstrate how the benefit of the denoising operation is dependent on the specification of the NO.
The presented analysis highlights the importance of objective IQ evaluation for DNN-based denoising technologies
and may suggest future avenues for improving their effectiveness in medical imaging applications.

The remaining of the paper is organized as follows. 
Section~\ref{sec:method} describes the necessary background on binary signal detection task, numerical observers, and image denoising.
The numerical studies and the results of the proposed evaluations of different denoising networks are provided in Sections~\ref{sec:numerical} and~\ref{sec:result}.
Finally, the article provides a discussion of the key findings in Sec.~\ref{sec:discussion}.

\section{Background}
\label{sec:method}

\subsection{Formulation of binary signal detection task}
\label{ssec:background}

A linear digital imaging system can be described as a continuous-to-discrete (C-D) mapping process \cite{barrett2013foundations}: 
\begin{equation}
\mathbf{g}=\mathcal{H}{f(\mathbf{r})+\mathbf{n}},   
\label{eq:imaging_system}
\end{equation}

\noindent where $\mathbf{g}\in\mathbb{R}^{N\times 1}$ is the measured image vector, 
$f(\mathbf{r})$ denotes the object function that is dependent on the coordinate $\mathbf{r}\in\mathbb{R}^{k\times 1}$, $k \ge 2$,
$\mathcal{H}$ denotes a linear imaging operator that maps $\mathbb{L}_{2}(\mathbb{R}^{k})$ 
to $\mathbb{R}^{N\times 1}$, 
and $\mathbf{n}\in\mathbb{R}^{N\times 1}$ denotes the measurement noise. 
When its spatial dependence is not important to highlight, $f(\mathbf{r})$ will be denoted as $\mathbf{f}$. 

A  binary signal detection task requires an observer 
to classify the measured image data $\mathbf{g}$ as satisfying either a signal-present hypothesis $H_1$ or a signal-absent hypothesis $H_0$.
These two hypotheses can be described as:
\begin{subequations}
	\label{eq:hypo}
	\begin{equation}
	H_0:\mathbf{g}=\mathcal{H}\mathbf{f_b}+\mathbf{n}=\mathbf{b+n}, 
	\end{equation}
	\begin{equation}
	H_1:\mathbf{g}=\mathcal{H}\mathbf{(f_b+f_s)}+\mathbf{n}=\mathbf{b+s+n},   
	\end{equation}   
\end{subequations}

\noindent where $\mathbf{f_s}$ and $\mathbf{f_b}$ denote the signal and background,
respectively, 
and $\mathbf{s}=\mathcal{H}\mathbf{f_s}$ and $\mathbf{b}=\mathcal{H}\mathbf{f_b}$ denote the signal and background images. 
For the case of a signal-known-exactly (SKE) and background-known-statistically (BKS) task,
$\mathbf{s}$ is known while $\mathbf{b}$ is a random vector.

To perform this task, a deterministic observer
computes a test statistic that maps the measured image $\mathbf{g}$ to a real-valued scalar variable that is compared to a
predetermined threshold $\tau$ to determine if $\mathbf{g}$ satisfies $H_0$ or $H_1$. 
By varying the threshold $\tau$, a ROC curve can be formed to quantify the trade-off between the false-positive fraction (FPF) and the true-positive fraction (TPF) \cite{barrett2013foundations}. The area under the ROC curve (AUC) can be subsequently calculated as a figure-of-merit (FOM) for signal detection performance.

\subsection{Numerical observers for IQ assessment}
\label{ssec:observers}

In preliminary assessments of medical imaging technologies, NOs have been  employed to quantify task-based measures of IQ for various image-based inferences\cite{barrett1993model}.
The NOs that are employed in this study to perform binary SKE/BKS signal detection tasks are described briefly below.

\subsubsection{Ideal Observer (IO) and CNN-based observer}
\label{ssec:CNN-IO}

The Bayesian Ideal Observer (IO) sets an upper limit of observer performance for signal detection tasks 
and has been advocated for use in optimizing medical imaging systems and data-acquisition designs~\cite{barrett2013foundations,metz1995toward,vennart1997icru,wagner1985unified,he2013model}.
The IO test statistic $t_{\text{IO}}(\textbf{g})$ is any monotonic transformation of the likelihood ratio $\Lambda_{\text{LR}}(\mathbf{g})$:
\begin{equation}
\label{eq:LR}
\Lambda_{\text{LR}}(\mathbf{g})=\frac{p(\mathbf{g}|H_1)}{p(\mathbf{g}|H_0)},
\end{equation}
\noindent where $p(\mathbf{g}|H_1)$ and $p(\mathbf{g}|H_0)$ are the conditional probability density functions that describe the measured data $\mathbf{g}$ under the hypotheses $H_1$ and $H_0$, respectively.
Equation~(\ref{eq:LR}) is analytically intractable, in general, and
 Markov-chain Monte Carlo (MCMC) techniques  have been proposed to approximate the  IO test statistic \cite{kupinski2003ideal}.
 In this study, an alternative method based on supervised learning is employed to approximate $\Lambda_{\text{LR}}(\mathbf{g})$. 
 Specifically, this will be accomplished by use
 of an appropriately designed CNN-based classifier as described elsewhere \cite{zhou2019approximating}.
 The resulting NO will be referred to as the \emph{CNN-IO observer}.
 
 Please note that when a CNN-based classifier is employed as a NO but it does not possess sufficient model capacity to accurately
 approximate $\Lambda_{\text{LR}}(\mathbf{g})$,
 it will simply be referred to as a \emph{CNN-based observer}. 
 Therefore, the CNN-based observer is, by definition, a sub-optimal observer.

\subsubsection{Hotelling Observer and regularized Hotelling observer} 

The Hotelling Observer (HO) is the IO that is restricted to employ test statistics that are linear functions
of the data \cite{barrett2013foundations}. The HO employs the Hotelling discriminant, which is the population equivalent of the Fisher linear discriminant, and is optimal among all linear observers in
the sense that it maximizes the signal-to-noise ratio of the test statistic~\cite{barrett2013foundations}.
The HO test statistic $t_{\text{HO}}(\mathbf{g})$ is defined as:
\begin{equation}
\label{eq:HO}
t_{\text{HO}}(\mathbf{g})=\mathbf{w}^{T}_{\text{HO}}\mathbf{g}
=(\mathbf{K_g}^{-1}\Delta\Bar{\mathbf{g}})^T\mathbf{g}, 
\end{equation}
\noindent where $\mathbf{w}^{T}_{\text{HO}}\in\mathbb{R}^{N}$ denotes the Hotelling template,
$\Delta\Bar{\mathbf{g}}\in\mathbb{R}^{N}$ denotes the difference between the ensemble mean of 
the measurements $\mathbf{g}$ under the two hypotheses $H_0$ and $H_1$, and
$\mathbf{K_g} \equiv \frac{1}{2}(\mathbf{K}_{0}(\mathbf{g})+\mathbf{K}_{1}(\mathbf{g}))$.
Here $\mathbf{K}_{0}(\mathbf{g}) \in \mathbb{R}^{{N}\times {N}}$ and $\mathbf{K}_{1}(\mathbf{g})\in \mathbb{R}^{{N}\times {N}}$ 
denote the covariance matrices of $\mathbf{g}$ under the two hypotheses $H_0$ and $H_1$. If a linear imaging system and a SKE signal detection task are considered, $\Delta\Bar{\mathbf{g}}=\mathbf{s}$. Note that the HO only employs first and second order statistical information about $\mathbf{g}$, whereas the IO requires full knowledge of the image data statistics.

In some cases, the covariance matrices 
$\mathbf{K}_{0}(\mathbf{g})$ and $\mathbf{K}_{1}(\mathbf{g})$ can be ill-conditioned and therefore the Hotelling template cannot be stably computed.
To address this, a regularized HO (RHO) can be employed that implements the test statistic $t_{\text{RHO}}(\mathbf{g})$:
\begin{equation}
	\label{eq:RHO}
	t_{\text{RHO}}(\mathbf{g})= \mathbf{w}^{T}_{\text{RHO}}\mathbf{g}  =(\mathbf{K}_{\lambda}^{+}	\Delta\Bar{\mathbf{g}})^T\mathbf{g},
\end{equation}
where $\mathbf{K}_{\lambda}$ represents a low-rank approximation of $\mathbf{K_g}$ that is formed by keeping only the singular values greater than  
$\lambda\sigma_{max}$.  Here,  $\mathbf{K}_{\lambda}^{+}$ is the Moore–Penrose inverse of $\mathbf{K}_{\lambda}$, 
$\lambda$ is a threshold for the singular values and $\sigma_{max}$ represents the largest singular value of $\mathbf{K_g}$, 
The value of $\lambda$ can be tuned on an independent set of data and the value that leads to the best RHO performance can be selected.

\subsubsection{Channelized Hotelling observer}

When the HO is employed with a channeling mechanism to reduce the dimensionality of the image data, a channelized HO (CHO) is formed.
When implemented with difference-of-Gaussian (DOG) channels and an internal noise mechansim, the CHO can be interpreted as an anthropomorphic observer~\cite{myers1987addition,abbey2000modeling,abbey2001human}.
Let $\mathbf{T}$ denote a channel matrix and $\mathbf{v\equiv Tg}$ the corresponding channelized image data. The CHO test statistic $t_{\text{CHO}}(\mathbf{g})$ is given by:
\begin{equation}
\label{eq:CHO}
t_{\text{CHO}}(\mathbf{g})=\left[(\mathbf{K_v+K_{int}})^{-1}\Delta\Bar{\mathbf{v}}\right]^T(\mathbf{v+v_{int}}),
\end{equation}
where $\mathbf{K_v}$ denotes the covariance matrix of the channelized data $\mathbf{v}$, 
$\mathbf{K_{int}}$ denotes the covariance matrix of the channel internal noise, 
and $\mathbf{v_{int}}$ is a noise vector sampled from the Gaussian distribution $\mathcal{N}(\mathbf{0,K_{int}})$. 
Based on previous studies \cite{abbey2001human}, in this work $\mathbf{K_{int}}$ will be defined as:
\begin{equation}
\label{eq:int_noise}
\mathbf{K_{int}}=\epsilon \cdot diag(\mathbf{K_v}),
\end{equation}
where $diag(\mathbf{K_v})$ represents a diagonal matrix with diagonal elements from $\mathbf{K_v}$ 
and $\epsilon$ is the internal noise level. The parameters of the DOG channels and the internal noise level 
employed in this study are described below in Sec.~\ref{ssec:numerical_observers}.

\subsubsection{Non-prewhitening matched filter (NPWMF)} 

The non-prewhitening matched filter (NPWMF) is a simple NO that utilizes only first-order statistical information~\cite{wagner1979application,burgess1994statistically}.
The NPWMF test statistic $t_{\text{NPWMF}}(\mathbf{g})$ is given by:
\begin{equation}
\label{eq:NPWMF}
t_{\text{NPWMF}}(\mathbf{g})=\Delta\Bar{\mathbf{g}}^T\mathbf{g},
\end{equation}
where $\Delta\Bar{\mathbf{g}}\in\mathbb{R}^{N}$ represents
the difference of the means of the ensemble of measured images $\mathbf{g}$ under the two hypotheses $H_0$ and $H_1$, respectively.
By design, the NPWMF will not be affected by changes to the second- and higher-order 
statistics of the image data.

\vspace{-0.1in}
\subsection{DNN-based image denoising}
\label{ssec:denoising-network}

Denoising methods based on DNNs hold significant potential for medical imaging applications~\cite{zhang2017beyond,yang2018low,manduca2009projection,li2014adaptive,lin2001improving,le2013denoising,gong2018pet,chen2017low,manjon2012new,jifara2019medical}. 
Due to their flexibility and ability to exploit image features,
many such denoising methods have been proposed based on CNNs. 
Given a noisy image $\mathbf{g}$, the action of a DNN-based denoising method can be described generically as:
\begin{equation}
\hat{\mathbf{g}}
= \mathcal{F}(\mathbf{g}, \mathbf{\Theta}),
\label{eq:denoising}
\end{equation}
where the mapping $\mathcal{F}$ denotes the DNN that is parameterized by the weight vector $\mathbf{\Theta}$
and $\hat{\mathbf{g}}$ denotes the estimated denoised image. 
{Depending on how the target data are defined when training the DNN, $\hat{\mathbf{g}}$ can be interpreted as an estimate of the noiseless $\mathbf{g}$ or an estimate of $\mathbf{g}$ that contains a reduced noise level.
When pre-training networks by use of simulated data,
the former approach has been commonly employed~\cite{manduca2009projection,li2014adaptive,lin2001improving,le2013denoising,gong2018pet,chen2017low,zhang2017beyond,manjon2012new,jifara2019medical,li2020mri}.}

In addition to CNN-based methods, a variety of other approaches, including residual learning~\cite{he2016deep}, have been employed for medical image denoising~\cite{gong2018pet,jifara2019medical}.
The performance of denoising networks has commonly been evaluated 
by use of traditional metrics such as structural similarity index metric (SSIM)~\cite{wang2004image} and peak signal-to-noise ratio (PSNR).

\section{Numerical Studies}
\label{sec:numerical}

Computer-simulation studies were conducted to objectively evaluate DNN-based denoising methods for SKE/BKS binary signal detection tasks.
Three different DNNs were investigated, which were trained on simulated image data.
The performances of the five different NOs reviewed in Sec.~\ref{ssec:observers} on the noisy  and denoised image data were analyzed under different conditions to gain insights into the potential impact of DNN-based denoising on signal detection.

\subsection{Simulated nuclear medicine images from a parallel-hole collimator imaging system}
\label{ssec:ske_bks} 

Planar scintigraphy images were simulated via an idealized linear parallel-hole collimator imaging system. 
The system was described by a linear C-D mapping $[\mathbf{g}]_m\equiv \int_V f(\mathbf{r})\,h_m(\mathbf{r})\, d\mathbf{r}$ that was specified by Gaussian point response functions~\cite{kupinski2003ideal}: 
\begin{equation}
h_m(\mathbf{r})={A_m}\exp\left[ -\frac{(\mathbf{r-r}_m)^T(\mathbf{r-r}_m)}{2w_{m}^2} \right], 
\label{eq:parallel-hole}
\end{equation}
where $[\mathbf{g}]_m$ denotes the $m^{th}$ component of $\mathbf{g}$, $V$ denotes the support of $f(\mathbf{r})$, 
and the amplitude $A_{m}=\frac{h}{2\pi w_{m}^2}$ 
with the height $h$ and width $w_m$.
The to-be-imaged objects $f(\mathbf{r})=f_b(\mathbf{r})+f_s(\mathbf{r})$ contained a random background and a superimposed deterministic signal in the signal present case. 
The random background $f_b(\mathbf{r})$ was specified by lumpy object model~\cite{kupinski2003ideal} as:
\begin{equation}
f_b(\mathbf{r})=\sum^{N_b}_{n=1}l(\mathbf{r-r}_n|a,w_b),
\label{eq:lumpy}
\end{equation}
where $N_b\sim P(\bar{N})$ denotes the number of the lumps with 
$P(\bar{N})$ denoting a Poisson distribution with the mean $\bar{N}$.
The lump function $l(\mathbf{r-r}_n|a,w_b)$ was modeled by a 2D Gaussian function with lump amplitude $a$ and lump width $w_b$:
\begin{equation}
l(\mathbf{r-r}_n|a,w_b)=a\exp(-\frac{(\mathbf{r-r}_n)^T(\mathbf{r-r}_n)}{2w^2_b}), 
\end{equation}
where $\mathbf{r}_n$ denotes the center location of the $n^{th}$ lump 
that was sampled from a uniform distribution over the spatial support of the image.
For the signal present cases, the signal corresponded to a Gaussian signal:
\begin{equation}
f_s(\mathbf{r})={A_s}\exp\left[ -\frac{(\mathbf{r-r}_s)^T(\mathbf{r-r}_s)}{2w_{s}^2} \right], 
\label{eq:sig_fun}
\end{equation}
where $A_{s}$ is the signal amplitude, $w_s$ is the signal width and $\mathbf{r_s}$ is the center of signal.
The images $\mathbf{s}=\mathcal{H}\mathbf{f_s}$ and $\mathbf{b}=\mathcal{H}\mathbf{f_b}$ are given by:
\begin{equation}
[\mathbf{s}]_m= \frac{{A_s}{h}{w_s^2}}{w_m^2+w_s^2}\exp\left[ -\frac{(\mathbf{r}_m-\mathbf{r}_s)^T(\mathbf{r}_m-\mathbf{r}_s)}{2(w_{m}^2+w_{s}^2)}\right],
\label{eq:sig_bks}
\end{equation}
and
\begin{equation}
[\mathbf{b}]_m= \frac{{a}{h}{w_b^2}}{w_m^2+w_b^2}\sum_{n=1}^{N_b}\exp\left[ -\frac{(\mathbf{r}_n-\mathbf{r}_m)^T(\mathbf{r}_n-\mathbf{r}_m)}{2(w_{m}^2+w_{b}^2)}\right].
\label{eq:b_img}
\end{equation}
The measurement noise $\mathbf{n}$ was described by an uncorrelated mixed Possion-Gaussian noise model. 
Details regarding the signal, background and noise are provided in Sec.~\ref{ssec:quant_linear} below.
Figure~\ref{fig:sample} shows an example of the signal and a noise free signal-present image  along with the corresponding noisy image data $\mathbf{g}$.

\begin{figure}[ht]
\vspace{-0.1in}
	\centering
	\subfigure[]
		{\includegraphics[width=0.151\textwidth]{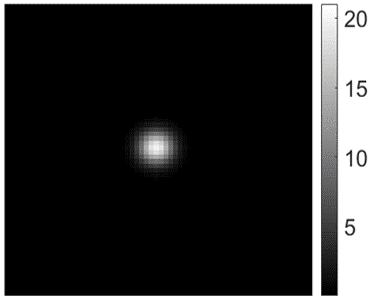}}
	\subfigure[]
		{\includegraphics[width=0.156\textwidth]{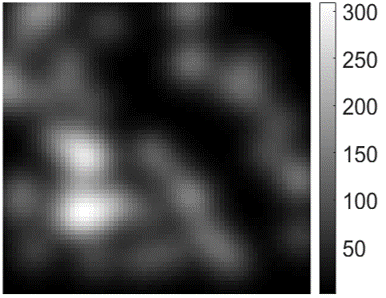}}
	\subfigure[]
	{\includegraphics[width=0.15\textwidth]{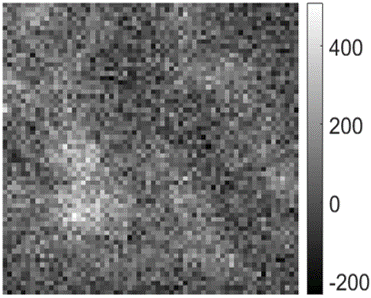}}
\vspace{-0.1in}
	\caption{These images are examples that depict (a) a possible signal $\mathbf{s}$, (b) a noise-free signal-present image $\mathbf{s+b}$, and (c) the corresponding noisy measurement $\mathbf{g}$. The dimensions of the images are $64\times64$. As described in the text, the signal amplitude was relatively small to emulate a situation where the detection task is challenging.}
	\label{fig:sample}
\vspace{-0.1in}
\end{figure}

The relatively simple image models employed in our study provided a means by which simulated image data could be computed and degraded in a clear and controlled way, without being influenced by unknown noise sources that could potential be present in clinically acquired images. 

\vspace{-0.1in}
\subsection{DNN-based denoising methods, training, and validation}
\label{ssec:denoising-network}

\begin{figure}[ht]
	\centering
	\subfigure[Linear convolutional layer-based denoising network]
	{
		\includegraphics[width=0.45\textwidth]{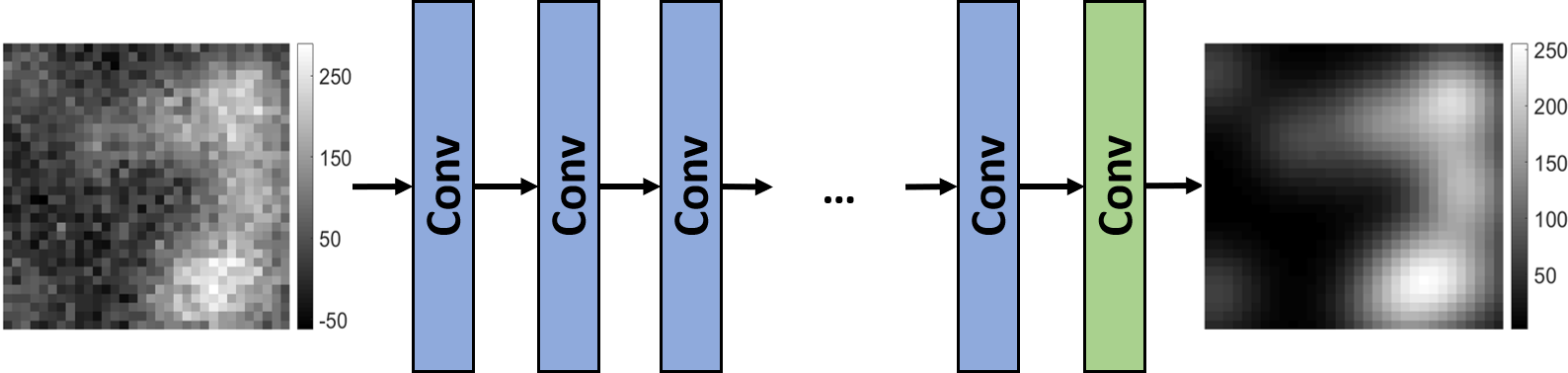}
		\label{fig:linear}
	}
	\subfigure[Nonlinear CNN architecture-based denoising network]
	{
		\includegraphics[width=0.45\textwidth]{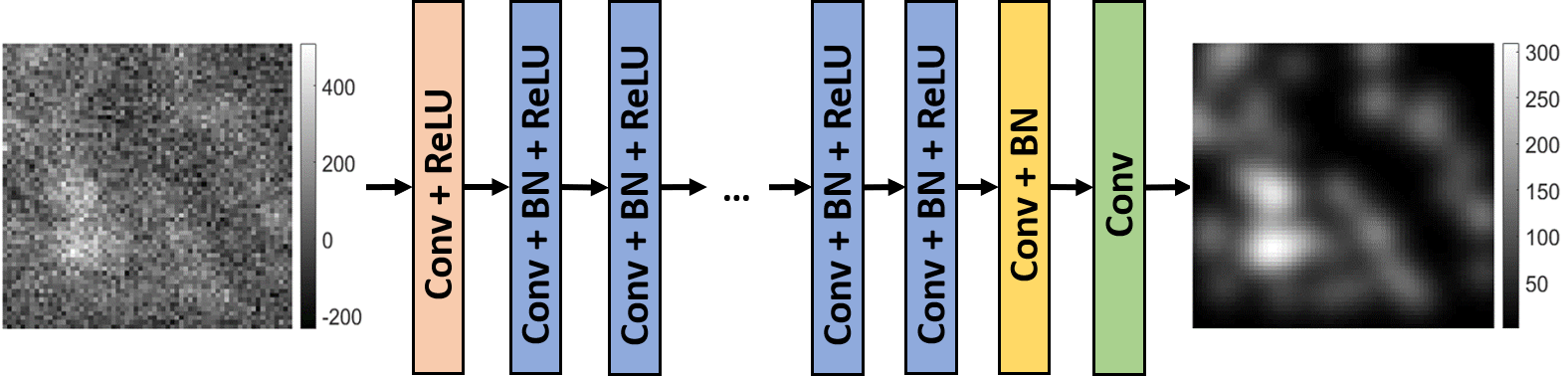}
		\label{fig:CNN_architecture}
	}
	\subfigure[Nonlinear ResNet architecture-based denoising network]
	{
		\includegraphics[width=0.455\textwidth]{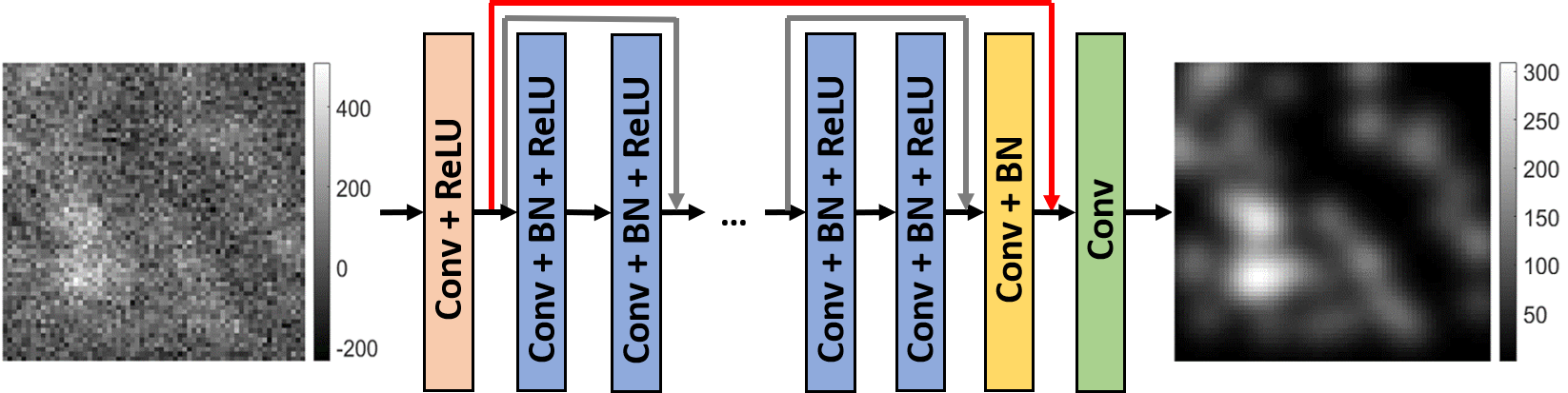}
		\label{fig:ResNet_architecture}
	}
	\caption{The three denoising networks evaluated in this study were based upon a (a) linear CNN, (b) non-linear CNN, and (c) non-linear ResNet denoising network, respectively. The dimensions of the input and output images are (a) 32$\times$32, (b) 64$\times$64, (c) 64$\times$64, respectively.}
	\vspace{-0.1in}
	\label{fig:denoising_networks}
\end{figure}

A simple linear denoising network 
and two nonlinear denoising networks with CNN-based or ResNet-based architectures were considered as three representative examples to be evaluated in this study.
Figure~\ref{fig:denoising_networks} shows the architectures of these three  networks, 
which are described next.

\subsubsection{Linear DNN-based denoising method}
\label{ssec:linear}

As depicted in Fig.~\ref{fig:linear}, the linear DNNs include only a collection
of $D$ linear convolutional layers.  Although such  networks will not achieve state-of-the art performance,
they are considered here because they permit the analytic propagation of covariance matrices, and hence Hotelling templates, through the different layers of the network. 
Therefore, preliminary insights into how DNNs perturb information relevant to binary signal detection tasks can be gained.
The network input was a noisy image $\mathbf{g}$ of dimension $32\times 32$ 
and the output was the estimated $\hat{\mathbf{g}}$ with the same dimensions.
In the first layer of the network, 32 filters of dimension $3\times3\times1$ were employed to generate $32$ feature maps. 
In each of the $2^{nd}$ to the {(${D}$-1)}$^{th}$ layers, 32 filters of dimension $3\times3\times32$ were employed. 
In the penultimate layer, a single filter of dimension $3\times3\times32$ was applied  to map the tensor-valued feature map to the scalar-valued output image. 

As described in Eqn.~(\ref{eq:imaging_system}), let $\mathcal{H}\mathbf{f}_j$ denote a given ground truth (noiseless) image corresponding to either a signal absent or signal present case and let $\mathbf{g}_j$ denote the corresponding measured noisy image. Here, the subscript $j$
has been added to index the objects and images.
Given the collection of paired training data
$\{(\mathbf{g}_j,\mathcal{H}\mathbf{f}_j)\}^J_{j=1}$, 
the linear network was trained
by minimizing the mean-square-error (MSE) loss function:
\begin{equation}
	\label{eq:MSE}
	\mathcal{L}_{\text{MSE}}(\mathbf{\Theta})=\frac{1}{J}\sum_{j=1}^J \Vert \mathcal{F}(\mathbf{g}_j;	\mathbf{\Theta})-\mathcal{H}\mathbf{f}_j\Vert_2^2.
\end{equation}
%

\noindent \subsubsection{Nonlinear CNN-based denoising network}
\label{ssec:CNN_denoising}

As depicted in Fig.~\ref{fig:CNN_architecture}, a traditional non-linear CNN architecture of depth $D$ was considered.
The network input was a noisy image $\mathbf{g}$ of dimension $64\times 64$ 
and the output was the estimated $\hat{\mathbf{g}}$ with the same dimensions.
The CNN contained four types of layers.
The first layer was a Conv+ReLU layer, in which 64 convolution filters of dimension $3\times3\times1$ were applied to generate 64 feature maps.
In each of the $2^{nd}$ to (${D}$-2)$^{th}$ Conv+BN+ReLU layers, 
64 convolution filters of dimension $3\times3\times64$ were employed 
and batch normalization was included between the convolution and ReLU operations. 
In the (${D}$-1)$^{th}$ Conv+BN layer, 64 convolution filters of dimension $3\times3\times64$ were employed and batch normalization was performed. 
In the last Conv layer, one single convolution filter of dimnension $3\times3\times64$ was employed to form the final denoised image of dimension $64 \times 64$.
The network was trained by use of the MSE-based loss function.

\subsubsection{Nonlinear ResNet-based denoising network}
\label{ssec:ResNet_denoising}

An alternative nonlinear denoising network based on a ResNet architecture \cite{he2016deep} was also investigated. 
As shown in Fig.~\ref{fig:ResNet_architecture},
the ResNet architecture employs shortcut connections (the so-called skip connections) 
between non-adjacent convolutional layers. 
This network design can better address the vanishing gradient issue \cite{he2016deep}, 
and allows for a deeper network with more convolutional layers.
In this study, skip connections were added every other layer, as depicted by the gray line in Fig.~\ref{fig:ResNet_architecture}. 
An additional skip connection, depicted as the brown line in Fig.~\ref{fig:ResNet_architecture}, was added 
to connect the output of the $1^{st}$ layer and the input of the  $D^{th}$ (i.e., last) layer.
Except for the skip connections, the network architecture was identical to that described above for the non-linear CNN.

Instead of using MSE-based loss, the perceptual loss was employed to train this network:
\begin{equation}
\label{eq:perceptual}
\mathcal{L}_{\text{Perceptual}}(\mathbf{\Theta})=\frac{1}{J}\sum_{j=1}^J \Vert \phi(\mathcal{F}(\mathbf{g}_j;\mathbf{\Theta}))-\phi(\mathcal{H}\mathbf{f}_j)\Vert_2^2,
\end{equation}
where $\phi(\cdot)$ represents a feature extraction operator. 
It has been observed that denoising networks trained by use of a perceptual loss function can be effective in reducing noise while retaining image details~\cite{gong2018pet}.

\subsubsection{Datasets and denoising network training details}
\label{ssec:network_training}

The standard convention of utilizing separate training/validation/testing datasets was adopted.
The training dataset included 10,000 noisy signal-present and 10,000 noisy signal-absent measurement images along with the corresponding noise-free target images.
The validation datatset included 200 signal-present images and 200 signal absent images and the corresponding noise-free 
target images.
Finally, the testing dataset comprised 10,000 signal-present images and 10,000 signal-absent noisy images.

These datasets were computed as follows. First,  lumpy background images, which were generated according to Eqn. (\ref{eq:b_img}), were employed as the noise-free signal-absent images. Then, a Gaussian signal was inserted to the background images to create noise-free images under the signal-present hypothesis. The signal was defined in Eqn. (\ref{eq:sig_bks}). Finally, mixed Poisson and Gaussian noise was added to the noise-free images under both hypotheses. The training, validation, or testing datasets were generated separately according to the steps described above. The statistical properties of these images varied between studies and are described  below.

All the denoising networks were trained on mini-batches at each iteration by use of the Adam optimizer~\cite{kingma2014adam} with a learning rate of 0.0001.
Each mini-batch contained 200 signal-present images and 200 signal absent images 
that were randomly selected from the training dataset.
The network model that possessed to the best  performance on the validation dataset was selected for use. 
Keras~\cite{chollet2015keras} was employed for implementing and training all networks on a single NVIDIA TITAN X GPU.

When training the nonlinear ResNet-based denoising network, 
the output before the first pooling layer from a pre-trained VGG19~\cite{simonyan2014very} network was employed as a feature extraction operator to compute the perceptual loss in Eqn. (\ref{eq:perceptual}). A similar feature extraction operator was utilized by Gong \etal~\cite{gong2018pet}.
The VGG19 network contained 16 convolutional layers, 5 max pooling layers, and 3 fully connected layers, 
and was trained by use of images from ImageNet~\cite{deng2009imagenet}. 
A total of 64 feature maps were extracted with spatial size $64 \times 64$ to compute the perceptual loss.

\subsection{Objective evaluation of denoising networks}
\label{ssec:quant_linear}

\subsubsection{Studies involving linear denoising networks}
\label{sssec:quant_linear}

A study was implemented to assess the performance of the RHO when acting on data corresponding to the outputs of different intermediate layers in the linear denoising network. In this way, the RHO performance could be observed as it propagates through the network.  The RHO was utilized because the resulting covariance matrices were generally ill-conditioned.

In the detection task, the signal defined in Eqn.~(\ref{eq:sig_fun}) was employed with $A_s=2.5$, $w_s=1$, and $\mathbf{r}_s=[16;16]^T$. 
The parameters of the lumpy background model defined in Eqn.~(\ref{eq:lumpy}) were $\bar{N}=15$, $a=5$, and $w_b=3$.
The dimensions of $\mathbf{s}$, $\mathbf{b}$, and $\mathbf{n}$ and $\mathbf{g}$ in Eqn.~(\ref{eq:hypo}) were $32\times32$. 
The assumed parameters of the imaging system defined in Eqn.~(\ref{eq:parallel-hole}) were $A_{m}=\frac{h}{2\pi w_{m}^2}$, $h=20$ and $w_m=2$.  
For the mixed Possion-Gaussian noise, the Gaussian noise was sampled from a Gaussian distribution with the mean 0 and the standard deviation 25.
Based on these settings,
the training/validation/testing datasets were established and the linear denoising networks with  depths varying from $D$=$2$ to $D$=$15$ were trained as described above in Sec.~\ref{ssec:linear}.  
Each network with different $D$ was trained separately to achieve the optimal performance based on the defined loss function. 

In order to compute the RHO acting on the tensor-valued feature data produced by each network layer, the covariance matrix $\mathbf{K}_{d}$ of the output data tensor of each layer needed to be estimated. Here, $d$ denotes a layer index. 
To accomplish this,
the tensor-valued data were vectorized and the associated covariance matrices corresponding to each layer were computed by propagating the covariance matrix $\mathbf{K_0}$ of the noisy input image through the network. 
Details regarding this procedure are provided in Sec.\textcolor{blue}{~1} of the Supplementary file.

\subsubsection{Studies involving the non-linear denoising networks}
\label{sssec:nonlinear}

A study was designed to investigate the performance of NOs
when acting on the original noisy measurement images and the corresponding denoised images produced by the non-linear CNN and ResNet-based networks. Several parameters of the simulated images and  denoising networks  were varied to gain insights into the potential impact of denoising on NO performance.

\begin{figure}[h]
	\centering
	\subfigure[Four examples of $\hat{\mathbf{g}}$ estimated by the CNN-based denoising networks]{
		\includegraphics[width=0.11\textwidth]{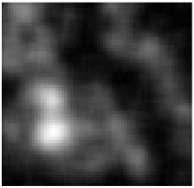}
		\includegraphics[width=0.11\textwidth]{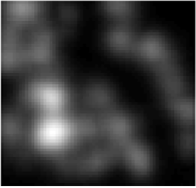}
		\includegraphics[width=0.11\textwidth]{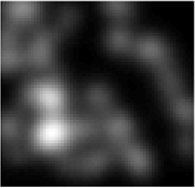}
		\includegraphics[width=0.11\textwidth]{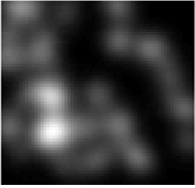}
		\includegraphics[width=0.0243\textwidth]{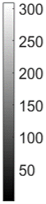}
	} 
	\subfigure[Four examples of $\hat{\mathbf{g}}$ estimated by the ResNet-based denoising networks]{
		\includegraphics[width=0.11\textwidth]{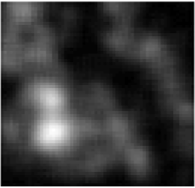}
		\includegraphics[width=0.11\textwidth]{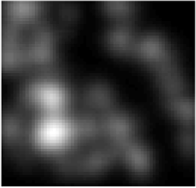}
		\includegraphics[width=0.11\textwidth]{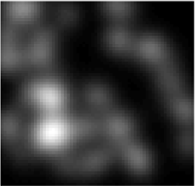}
		\includegraphics[width=0.11\textwidth]{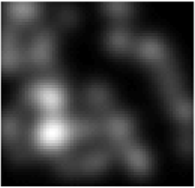}	
		\includegraphics[width=0.0243\textwidth]{clean_colorbar.png}
	}
	\caption{The images, from left to right, in each row represent the denoised estimates $\hat{\mathbf{g}}$ obtained by use of a) the CNN-based and b) the ResNet-based non-linear networks with varied \{3, 7, 11, 13\} layers, respectively.
	The related noise-free signal-present target image $\mathcal{H}(\mathbf{f_s}+\mathbf{f_b})$ and the original noisy image $\mathbf{g}$ were the second and third images shown in Fig.~\ref{fig:sample}. The dimensions of the images are $64\times64$.}
	\label{fig:nonlinear_denoising}
\end{figure}

For the considered detection tasks, the signals, the lumpy object model, and the parallel-hole collimator imaging system were defined as in Sec.~\ref{sssec:quant_linear} but with different parameter settings. 
The signal possessed an amplitude $A_s=3$,  width $w_s=\sqrt{2}$, and  center location $\mathbf{r}_s=[32;32]^T$. 
The parameters of the lumpy background model defined in Eqn.~(\ref{eq:lumpy}) were $\bar{N}=50$, $a=5$, and $w_b=3$. 
The dimensions of $\mathbf{s}$, $\mathbf{b}$, 
and $\mathbf{n}$ in Eqn.~(\ref{eq:hypo}) were $64\times64$. 
The parallel-hole collimator imaging system was specified as $A_{m}=\frac{h}{2\pi w_{m}^2}$, $h=20$ and $w_m=2$.
The standard deviation of Gaussian noise was set to 75.
{Based on these settings, the training/validation/testing datasets were established and   nonlinear denoising networks of depth  $D=\{3, 5, 7, 9, 11, 13\}$ were trained as described above in Sec.~\ref{ssec:CNN_denoising}.}
Examples of denoised images $\hat{\mathbf{g}}$ produced by use of the CNN-based and ResNet-based denoising networks of different depths $D$ are shown in Fig.~\ref{fig:nonlinear_denoising}.
This study was also repeated for the case where low-noise, instead of
noiseless, target images, were employed for training. 
Those studies are presented in Sec.\textcolor{blue}{~2} of the Supplementary file.

Finally, the impact of the signal size on the performance of the RHO was investigated.
Signals of width $w_s=\{1,\sqrt{2},2,2.5,3\}$ were considered. 
All other parameters were kept the same as that described above.

\begin{table*}[h]
	\centering
	\caption{RHO signal detection performance propagation through linear CNN-based denoising network with \{3, 5, 7, 9, 11, 13, 15\} layers were demonstrated by use of AUC values.
			The standard error of each AUC value was the same of 0.003.}
	\begin{tabular}{ c c c c c c c c c}
		\hline\hline
		   \multicolumn{9}{c}{\textbf{RHO detection performance (AUC value) at the output of different layers of the linear denoising network}} \\
		   \hline
	\textbf{Noisy}	& \textbf{Layer}	          & \multicolumn{7}{c}{\textbf{The denoising network with different layers}}\\
	\textbf{measurements}	& \textbf{Index} & \textbf{3 layers} & \textbf{5 layers} & \textbf{7 layers} & \textbf{9 layers} & \textbf{11 layers} & \textbf{13 layers} & \textbf{15 layers} \\
	 \multirow{7}{*}{0.6376}   &	3 & 0.6376 & 0.6376 & 0.6376 & 0.6376 & 0.6376 & 0.6376 & 0.6376 \\
		                                   & 5 &        & 0.6372 & 0.6376 & 0.6376 & 0.6376 & 0.6376 & 0.6376 \\
		                                   & 7 &        &        & 0.6316 & 0.6376 & 0.6376 & 0.6376 & 0.6376 \\
		                                   & 9 &        &        &        & 0.6283 & 0.6376 & 0.6376 & 0.6376 \\
                                              &	11 &        &        &        &        & 0.6213 & 0.6376 & 0.6376 \\
							&	13 &        &        &        &        &        & 0.6188 & 0.6376 \\
							&	15 &        &        &        &        &        &        & 0.6158 \\
		\hline\hline
	\end{tabular}
	\label{tab:RHO_each}
\end{table*}

\subsubsection{Observer performance evaluation metrics}
\label{sssec:metrics}

To evaluate the performance of the NOs, ROC analysis was conducted and
AUC values were computed and employed as a figure-of-merit.
The  ROC curves were fit by use of the Metz-ROC software~\cite{metz} that employs the proper binormal model~\cite{pesce2007reliable}.
The error bars of the  AUC values were estimated as well.
Detection efficiencies for a given NO and denoising method were defined as
\begin{equation}
	\label{eq:detectability}
	\mathit{e}\equiv\frac{\text{AUC}_\text{{denoised}}}{\text{AUC}_\text{{noisy}}},
\end{equation}
where $\text{AUC}_\text{{denoised}}$ and $\text{AUC}_\text{{noisy}}$ denote the AUC values corresponding to a NO acting on the denoised and original noisy image data, respectively.
The detection efficiency quantifies the impact of
the denoising operation on the performance of the NO.
It should be noted that    this definition is different from that employed elsewhere in the literature, where detection efficiency is typically referenced to an IO \cite{park2005efficiency}.  
As such, it is possible that $e>1$ when the IO is not employed.
The denoised images were also assessed by use of RMSE and SSIM.

\subsubsection{Numerical observer computation}
\label{ssec:numerical_observers}

The CNN-IO was employed to approximate the IO test statistic~\cite{zhou2019approximating}. 
Details regarding the implementation of the CNN-IO and CNN-based observers are provided in Sec.\textcolor{blue}{~5} of the Supplementary file.

For computing the HO and RHO test statistics, the covariance matrix $\mathbf{K_g}$ need to be estimated.
For use in evaluating the linear denoising networks, {the covariance matrix decomposition method~\cite{barrett2013foundations,zhou2019approximating}} was initially employed to estimate the covariance matrix of the original noisy images. To estimate the covariance matrix of the  
background images, 100,000 signal-present and 100,000 signal-absent noiseless images were utilized.
Subsequently, to examine how task-performance propagates through the networks, the covariance matrices corresponding to the
vectorized feature tensors at each network layer were 
  computed by use of the propagation strategy described in Sec. 1 of the Supplemental file.
For evaluating the nonlinear denoising networks, the covariance matrices corresponding to both the noisy and denoised images were empirically estimated by use of 100,000 signal-present and 100,000 signal-absent images.

{When computing the RHO test statistic, the threshold parameter $\lambda$ in Eqn.~(\ref{eq:RHO}) was swept from $1e-3$ to $1e-7$ and the corresponding detection performance was estimated based on a separate validation dataset including 2,000 signal-present images and 2,000 signal-absent images.} The value which led to the best RHO detection performance was selected.
{The RHO with selected parameter was then applied to the testing dataset described below and the corresponding observer performance was estimated.}
The NPWMF template was established by use of the same training data as employed to establish the RHO. 

For computing the CHO test statistic, 2,000 signal-present and 2,000 signal-absent images were utilized to estimate the channelized covariance matrix. 
A  set of 10 DOG channels~\cite{abbey2001human} was employed with channel parameters $\sigma_0=0.005$, $\alpha=1.4$, and $Q=1.67$. 
The internal noise level $\epsilon$ was 2.5, which was the same value employed by Abbey~\etal \cite{abbey2001human}.

The performance of the NOs on the original noisy images was evaluated by use of a testing dataset with 10,000 signal-present noisy images and 10,000 signal-absent noisy images that was described above in Sec.~\ref{ssec:network_training}.
Subseqently, the performance of the NOs was assessed by use of the denoised testing images.

\section{Results}
\label{sec:result}

\subsection{Propagation of task-based information through a linear denoising network}
\label{sec:result_a}

The performance of the RHO acting on the noisy test data and  on data corresponding to the outputs of different intermediate layers in the linear denoising network is summarized in Table~\ref{tab:RHO_each}.
The covariance matrices needed to compute the RHO test statistic
corresponding to the output of each network layer  were calculated by use of the propagation strategy described in Sec.\textcolor{blue}{~1} of the supplementary file.
With the exception of the network with three layers, the RHO performance on the denoised images was lower than on the original noisy images, and the performance decreases more on the image denoised by deeper networks.

\begin{figure}[ht]
	\centering
	\includegraphics[width=0.45\textwidth]{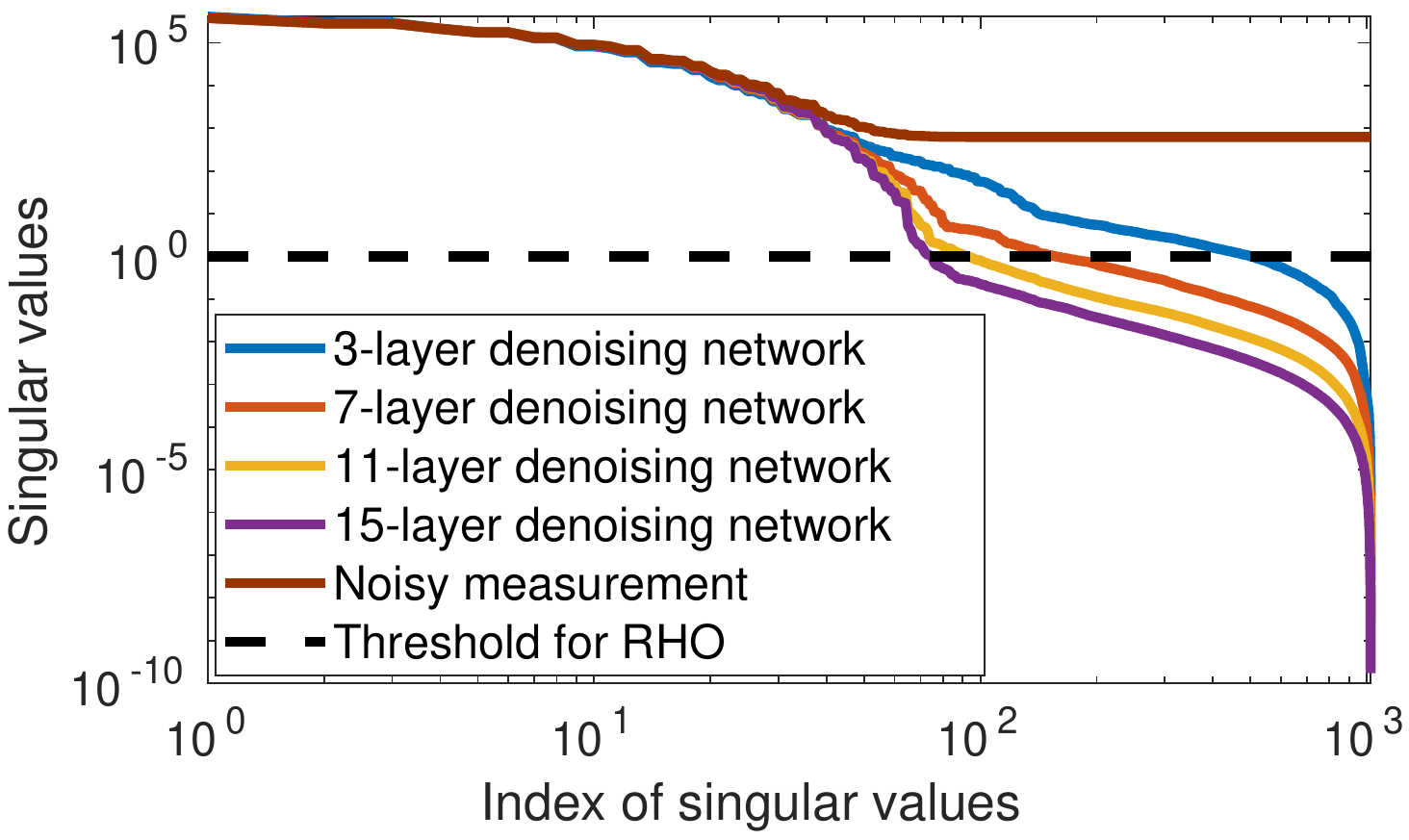}
	\caption{The singular value spectra of the covariance matrices corresponding to the original noisy images and the images denoised by use of the linear denoising networks with depths of $\{3,7,11,15\}$ were demonstrated.
	}
	\label{fig:linear_noisy_svd}
	\vspace{-0.1in}
\end{figure}

To gain insights into this behavior, the singular value spectra of the covariance matrices estimated from the original noisy images and   from the images denoised by networks with varied depths were examined. 
The results, shown in Fig.~\ref{fig:linear_noisy_svd}, reveal that the spectra corresponding to the denoised images
decay more rapidly than that corresponding to the original noisy image.  Additionally, the spectra corresponding
to the denoised images decayed more rapidly as the denoising network became deeper.
Accordingly, the number of singular values that exceeded the value of the threshold $\lambda\sigma_{max}$ that specified the RHO via Eqn.~(\ref{eq:RHO}) decreased as the network depth increased. 
This resulted in  the RHO performance to degrade as the network depth increased.

\begin{figure}[ht]
	\centering
	\includegraphics[width=0.45\textwidth]{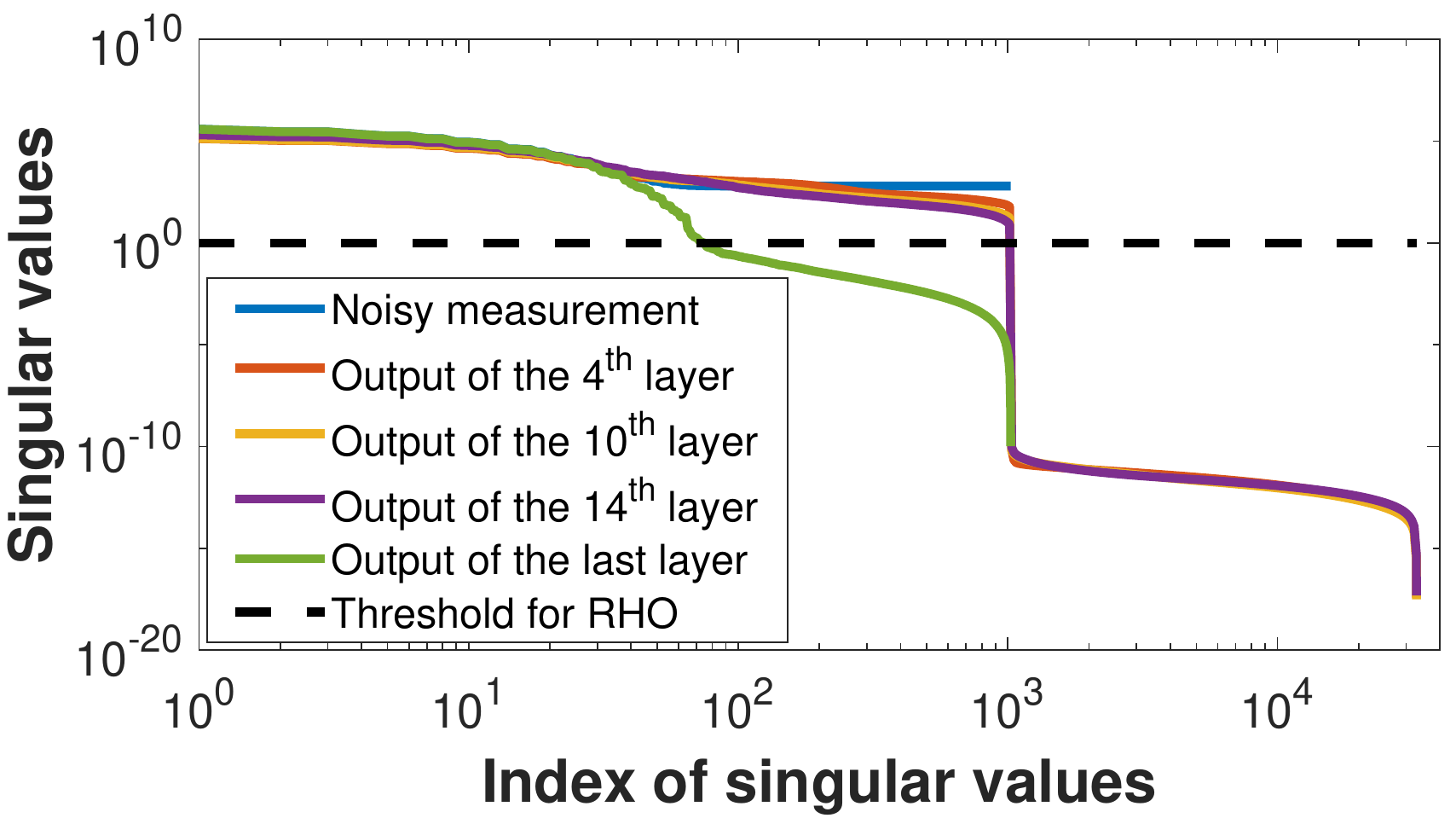}
	\caption{The singular value spectra of covariance matrices corresponding to the original noisy images and the outputs of different layers in a linear CNN denoising network with the depth $D=15$ were illustrated.
	}
	\label{fig:linear_svd}
	\vspace{-0.1in}
\end{figure}

\begin{figure}[h]
	\centering
	\includegraphics[width=0.46\textwidth]{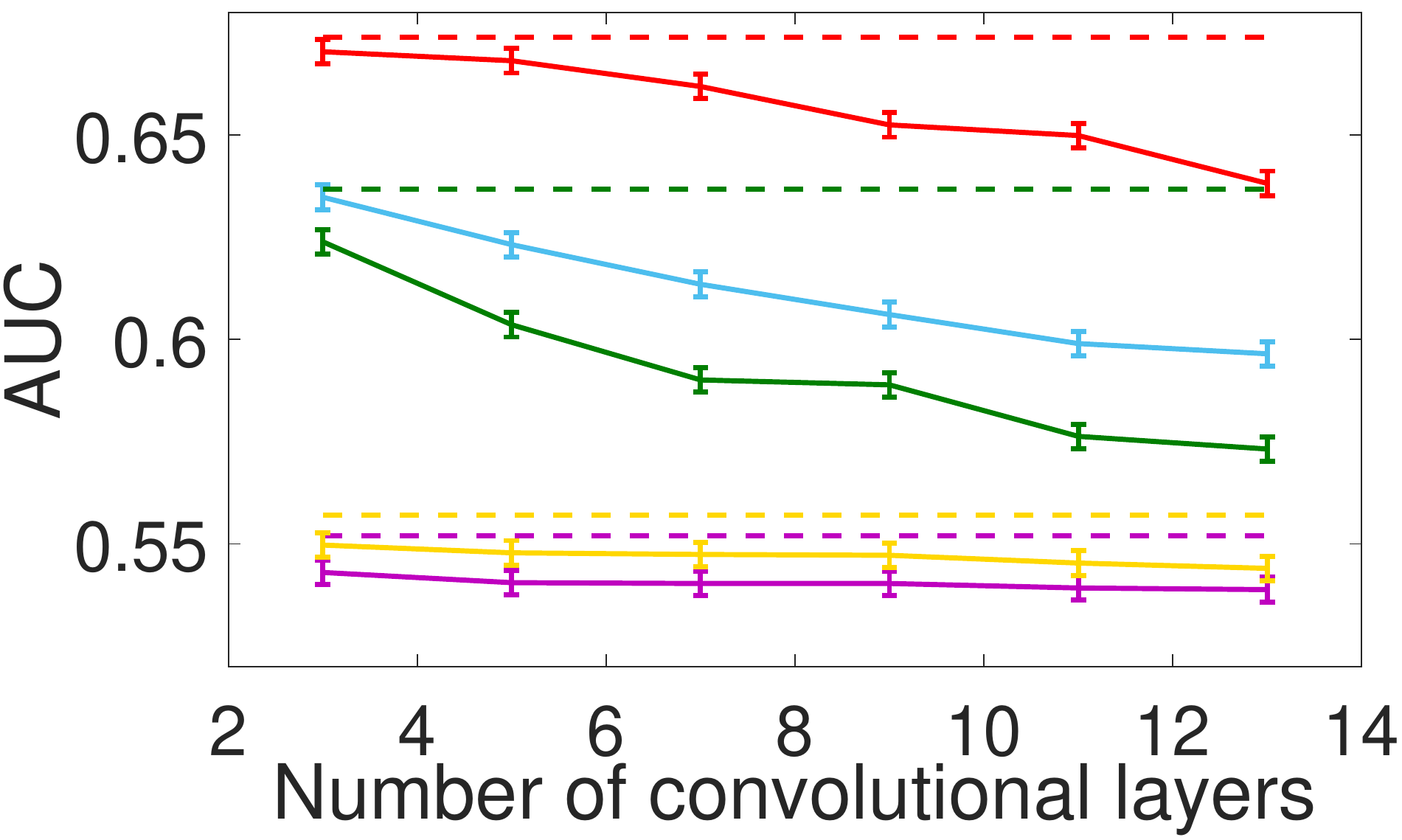}
\hspace{0.1in}
	\includegraphics[width=0.46\textwidth]{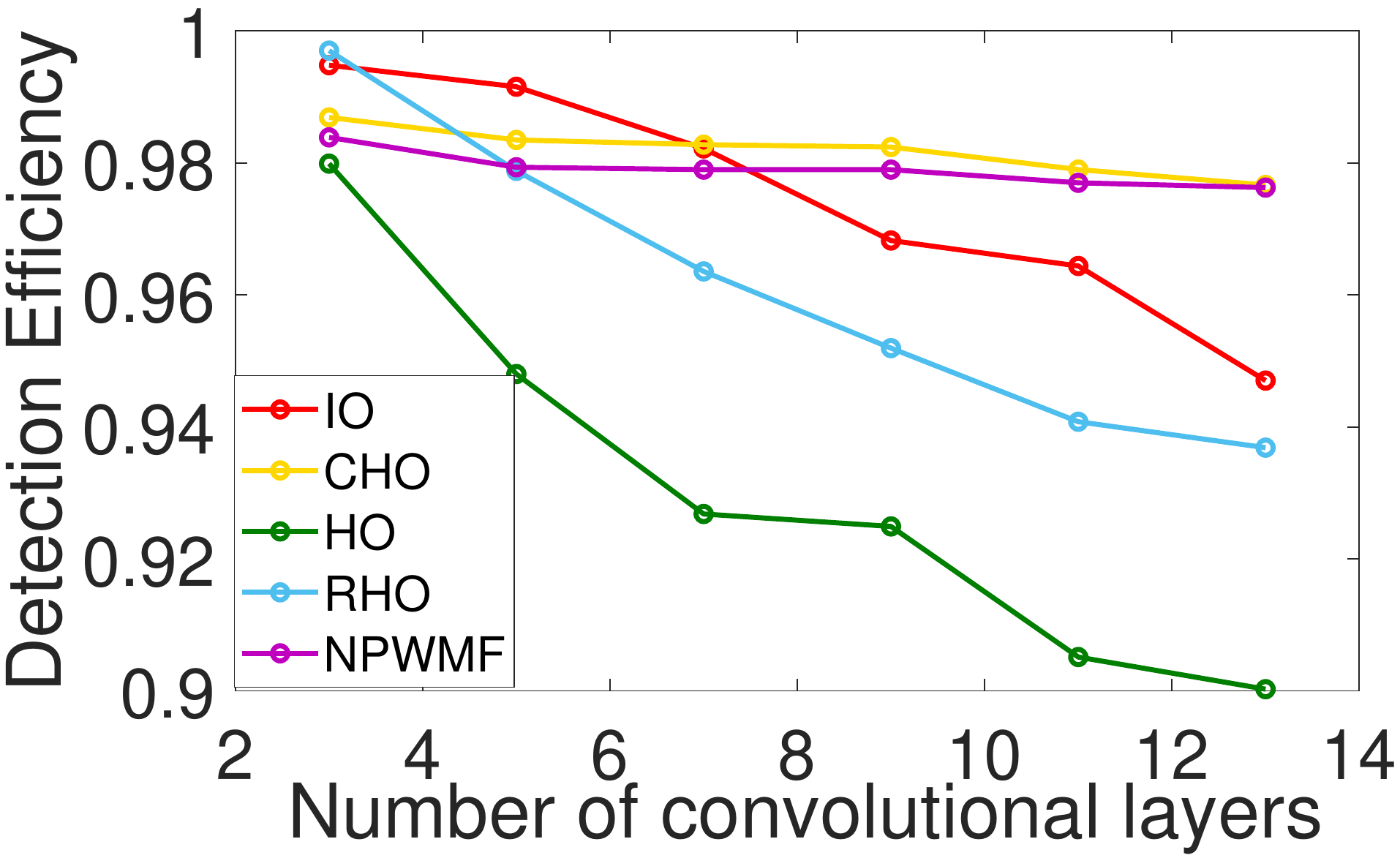}
	\caption{The relationships between AUC (top figure) and detection efficiency (bottom figure) and the depth (the number of convolutional layers) of a CNN-based non-linear denoising method when different NOs are employed were quantified.
 	The two figures share the same legend that is displayed in the bottom figure.
 	The dashed lines in the upper figure 
 	depict the performances of the NOs on the original noisy images.}
\label{fig:results1-CNN}
\end{figure}

\begin{table}[ht]
	\centering
	\caption{The RMSE and SSIM values associated with noisy images and the output images of two different nonlinear denoising networks were compared.}
	\begin{tabular}{c|c| c| c |c }
    \hline\hline
    \textbf{}	 &  \multicolumn{4}{c}{\textbf{Measurement Metrics }} \\	
    \textbf{Denoising}	 & \multicolumn{2}{c}{\text{CNN + MSE }} & \multicolumn{2}{c}{\text{ResNet + Perceptual }} \\
    \textbf{networks}	 &  RMSE & SSIM &  RMSE & SSIM  \\
    \hline
    Noise image & 75.4161 & 0.3663 & 75.4161   & 0.3663\\
    3 layers  		& 13.1478 & 0.9370  & 13.3280  &  0.9337\\
    5 layers   	& 12.1819 & 0.9469  & 12.3120     &  0.9463\\
    7 layers   	& 11.5499 & 0.9526 & 11.6433  & 0.9519\\
    9 layers   	& 11.4584 & 0.9535  & 11.4340  &  0.9540 \\
    11 layers 	& 11.4563 &  0.9536 & 11.3016  & 0.9549\\
    13 layers 	& 11.4548 & 0.9537  & 11.2556  &  0.9555 \\
	\hline\hline
	\end{tabular}
	\label{tab:physical-CNN}
	\vspace{-0.1in}
\end{table}

\begin{figure}[h]
	\centering
	\includegraphics[width=0.45\textwidth]{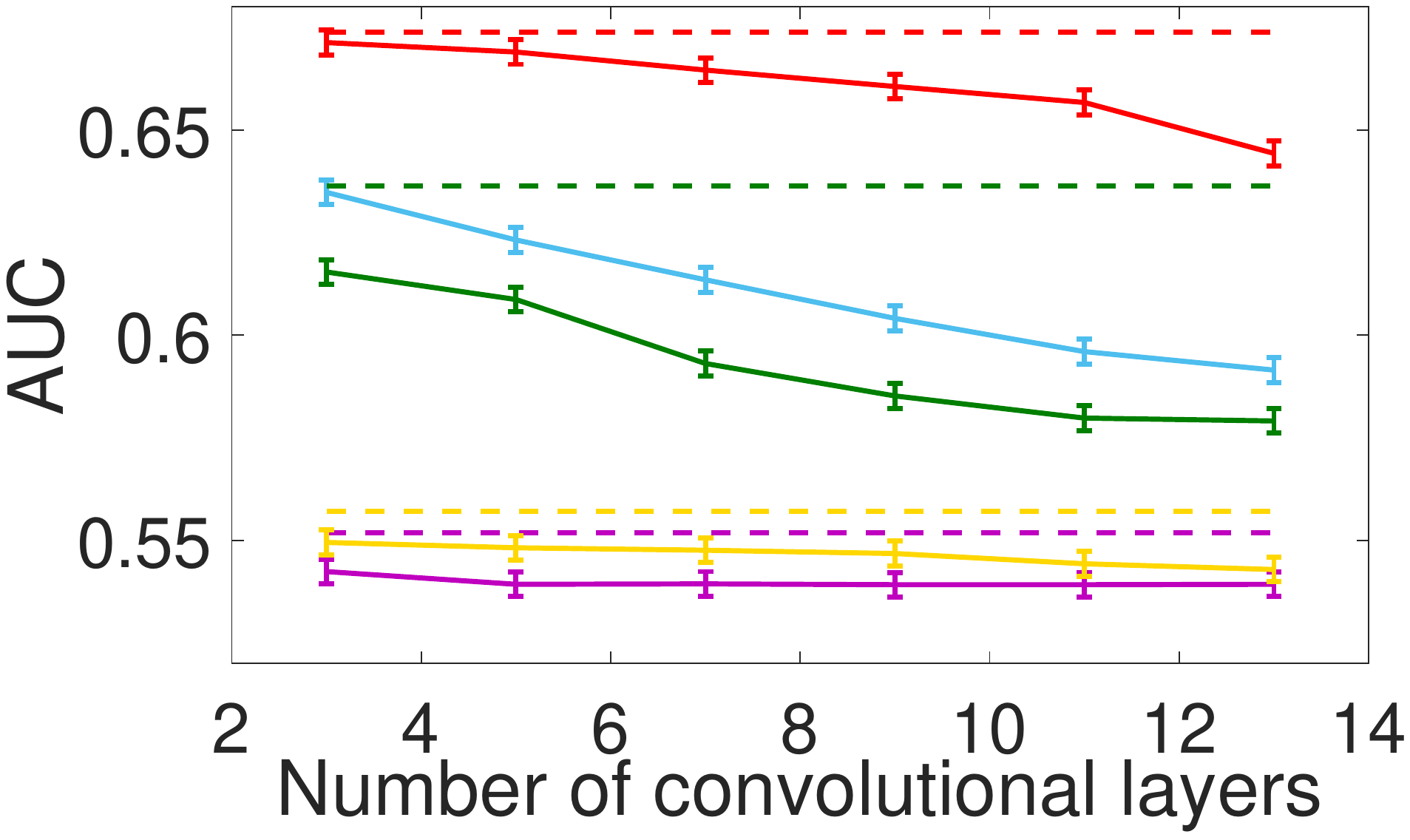}
	\hspace{0.1in}
	\includegraphics[width=0.45\textwidth]{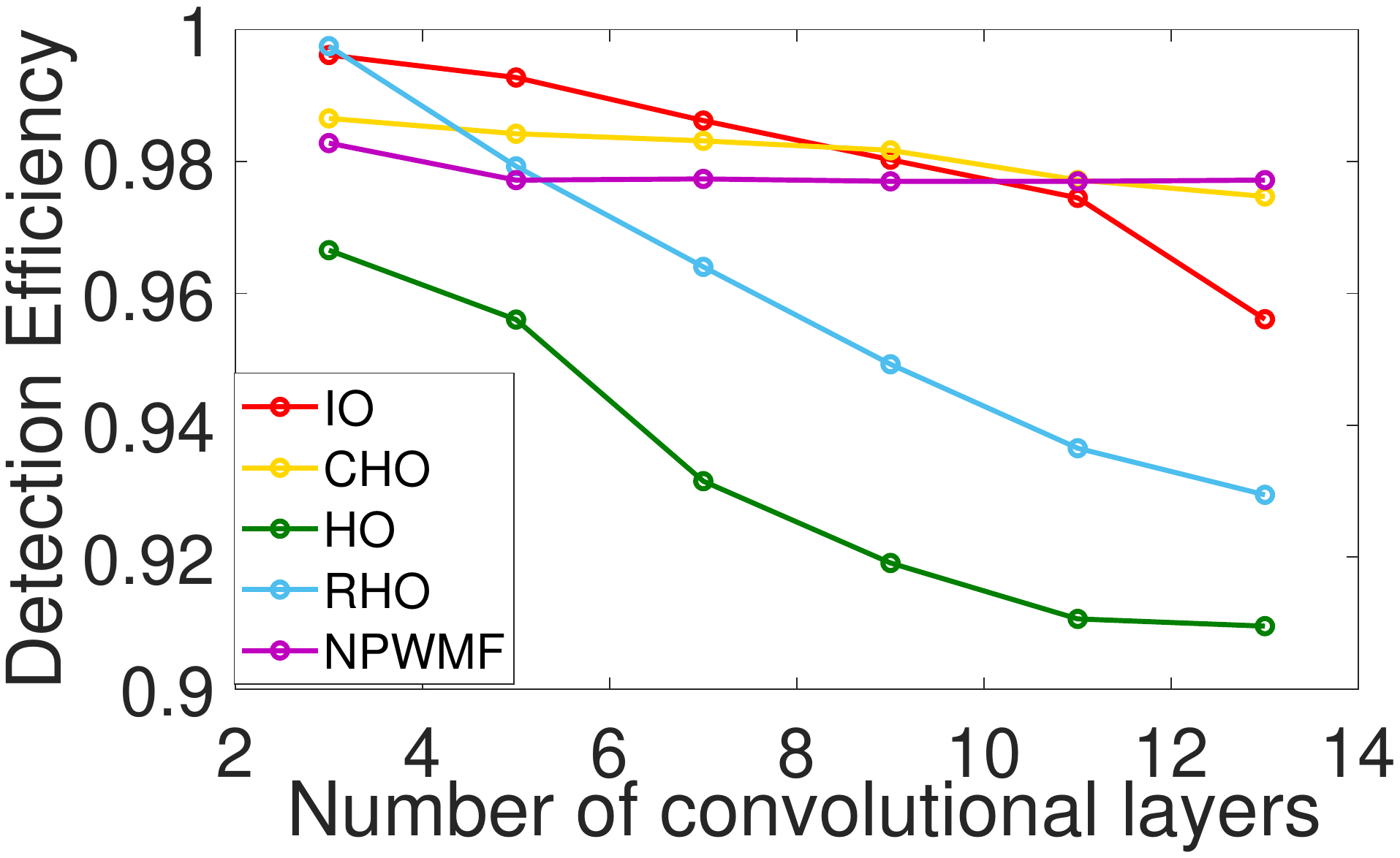}  
	\caption{The relationship between NO performance and the depth (the number of convolutional layers) of the ResNet-based non-linear denoising networks was quantified. 
 	The two figures on each panel share the same legend.
 	The dashed lines in the upper figure represent the performance of the NOs on the noisy images.}
 	\vspace{-0.1in}
\label{fig:results1-ResNet}
\end{figure}

The propagation of RHO performance through the networks is summarized in Table~\ref{tab:RHO_each}. It was
observed that the RHO performance on data produced by the intermediate denoising network layers remained approximately constant until the last layer, at which point it decreased. 
It should be noted that the last layer of the denoising network transforms a high-dimensional feature tensor to the denoised output image.  
This operation possesses   a null space and is therefore non-invertible.
The drop in RHO performance at the last layer suggests that
some of the features that were important to task-performance
 resided in the null
space of the learned transformation.

To understand why NO performance remained constant
until the last layer, the singular value spectra of the covariance matrices estimated from the original noisy images and the feature tensors corresponding to intermediate layers of the denoising network were further analyzed for the case of the network of depth $D=15$.
The results, shown in Fig.~\ref{fig:linear_svd}, reveal
that the spectra  corresponding to the intermediate layers were similar to that corresponding to the
original noisy images.
Accordingly, the number of singular values that exceeded the value of the threshold $\lambda\sigma_{max}$ that specified the RHO via Eqn.~(\ref{eq:RHO}) at different intermediate network layers remained constant as the network depth increased. 
This resulted in  the RHO performance remaining fixed as the network depth increased, until the last layer was reached as
discussed above.

\subsection{Impact of denoising network depth}

\subsubsection{Performance changes}
\label{ssec:nonlinear-performace}

The impact of depth of the non-linear CNN and ResNet networks on the NO performance as measured by AUC and detection efficiency is shown in Fig.~\ref{fig:results1-CNN} and Fig.~\ref{fig:results1-ResNet}.
For all cases, it was observed that the performance of the NOs on the original noisy images was higher than   on the  denoised images.  
The performance of the CNN-IO, HO and RHO on the denoised images  decreased as the 
depth of the denoising networks increased. 
Contrarily, the performance of CHO and NPWMF on the denoised images was relatively insensitive
to the depth of the denoising networks. These observations suggest that the second- and potentially higher-order
statistical properties of the images were degraded by the denoising networks; this is confirmed below in Fig.~\ref{fig:singular_values}.
The quality of the denoised images as measured by RMSE and SSIM values for the networks of
varying depth  are shown in Table~\ref{tab:physical-CNN}. As expected, these metrics improved
as the depth of the denoising networks increased.  
These results confirm that
objective measures of IQ based on signal detection performance can show conflicting
trends as compared to traditional metrics when comparing different denoising networks.

\subsubsection{Changes in covariance matrix induced by denoising}

\begin{figure}[ht]
	\centering
	\includegraphics[width=0.45\textwidth]{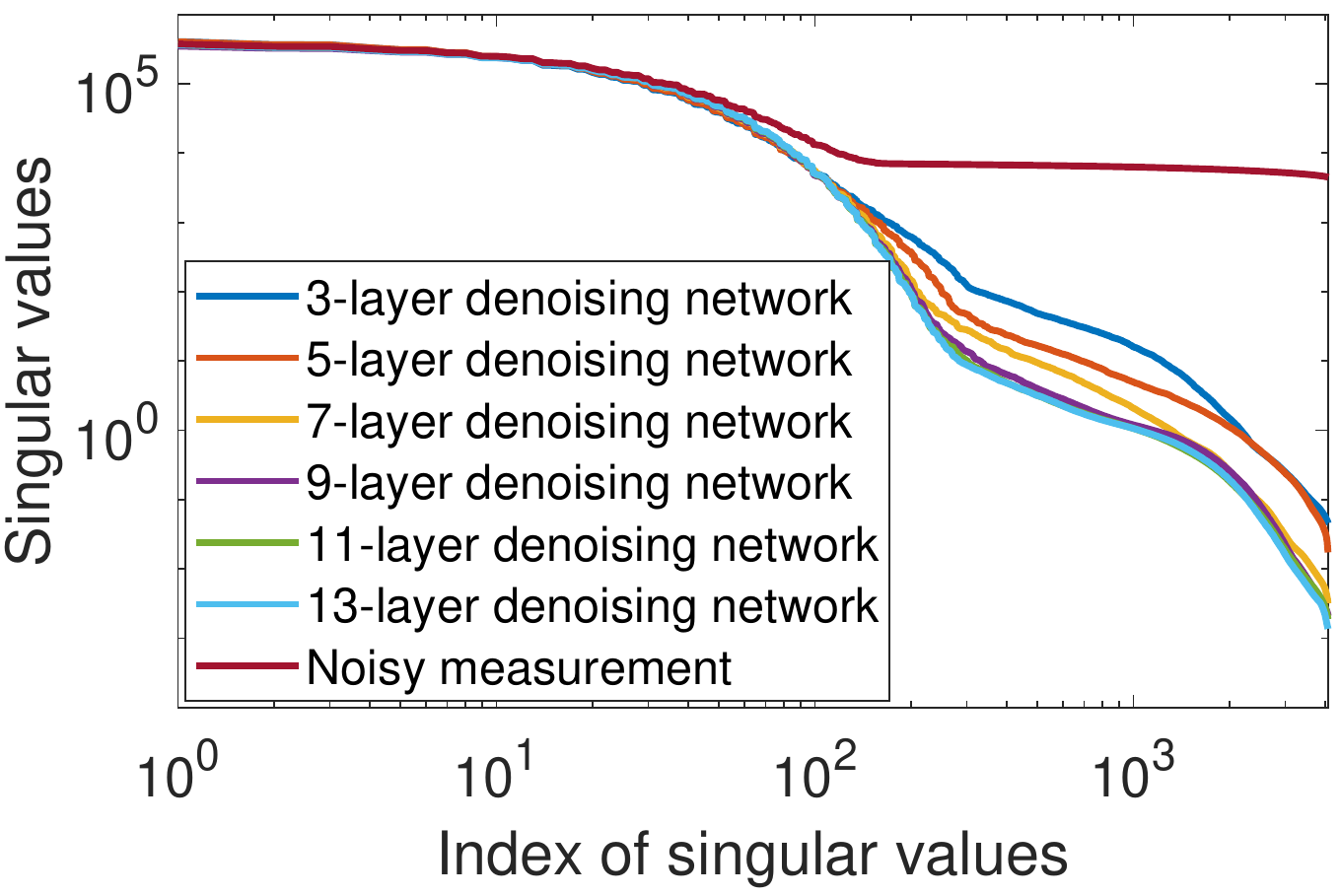}
	\caption{The singular values of the covariance matrices from noisy images and images denoised by CNN-based non-linear denoising networks with $\{3,5,7,9,11,13\}$ convolutional layers were compared, respectively. The denoising operation changes the structure of data covariance matrix. The changes are more obvious for deeper networks.
	}
	\vspace{-0.1in}
	\label{fig:singular_values}
\end{figure}

The degradation of HO performance was further analyzed by computing the SVD of
the covariance matrices corresponding to the images denoised by use of the CNN-based method.
The results, shown in Fig.~\ref{fig:singular_values}, reveal that the covariance matrix corresponding
to the denoised images was ill-conditioned, while that corresponding to the original noisy images was well-conditions.
Moreover, the singular value spectra tended to decrease more rapidly as the depth of the denoising
network was increased. Although not shown, similar observations were made in the case of the ResNet-based denoising
method. These results confirm that the denoising networks changed the second-order statistical properties of
the denoised images.  As mentioned above, the performance of the NPWMF observer, which uses only first-order statistical information, was not strongly degraded by denoising.  Together, these observations support the
assertion that the reduction in performance of the NOs that were sensitive to second- and higher-order image statistics was caused by the the changes in these properties induced by the denoising operation.

\subsection{Detection efficiency vs. signal size}
\label{ssec:sig_size}

\begin{figure}[ht]
	\centering
	\includegraphics[width=0.45\textwidth]{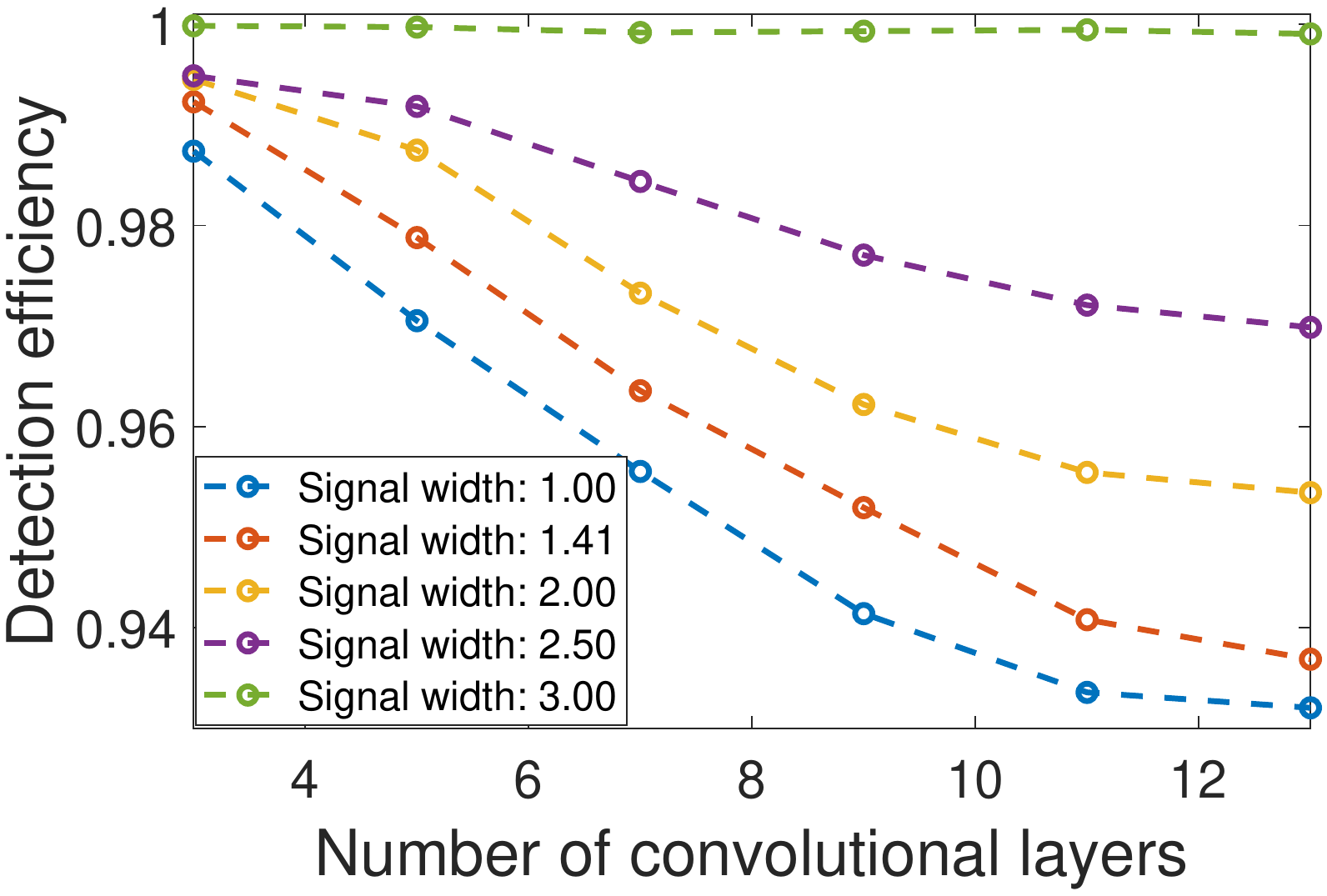}
	\caption{The relationships between signal size and RHO detection efficiency were quantified. 
		Here, the CNN-based denoising method was employed with an MSE loss and the network depth
		was varied: $D=\{3, 5, 7, 9, 11, 13\}$. 
		Detection efficiency   reduced more rapidly as a function of network depth when the signal size was reduced.
		}
	\label{fig:sig_size}
	\vspace{-0.1in}
\end{figure}

The impact of signal size on RHO detection efficiency
is shown in Fig.~\ref{fig:sig_size}.
Here, the width $w_s$ of the Gaussian signal in Eqn.~(\ref{eq:sig_fun}) took on the values: 
$\{1,\sqrt{2},2,2.5,3\}$.
It was observed that, for each signal size, the detection efficiency was reduced as the denoising network
depth increased.
Additionally, the detection efficiency   reduced more rapidly as a function of network depth for smaller signal sizes as compared to larger ones.
Specifically, for $w_s$ = 3, there was no
statistically significant decrease in detection efficiency as the denoising network depth increased. Moreover, the detection efficiency was close to one. This is due to the relatively large
size of the signal and use of an MSE loss function to train the denoising network.
An MSE loss function treats every pixel in an image equally and therefore a large signal
contributes more than a small one during the network training (i.e., more task-specific
information is potentially preserved).

\subsection{Situations where denoising improved detection performance}

CNN-based observers of varying depths were employed
 to demonstrate conditions under which the CNN- and ResNet-based denoising methods
 could improve signal detection performance. 
Detection performance was assessed on the original noisy images and the outputs of the two denoising networks with the depth of $\{3, 9, 11\}$, respectively. 
The evaluated CNN-based observers for this study were set with \{1, 2, 4, 6, 8, 10\} convolutional layers, respectively. It should be noted that the CNN-based observer with 10 layers coincided with the CNN-IO, and therefore approximated the IO for this task.

\begin{figure}[ht]
	\centering
	\includegraphics[width=0.46\textwidth]{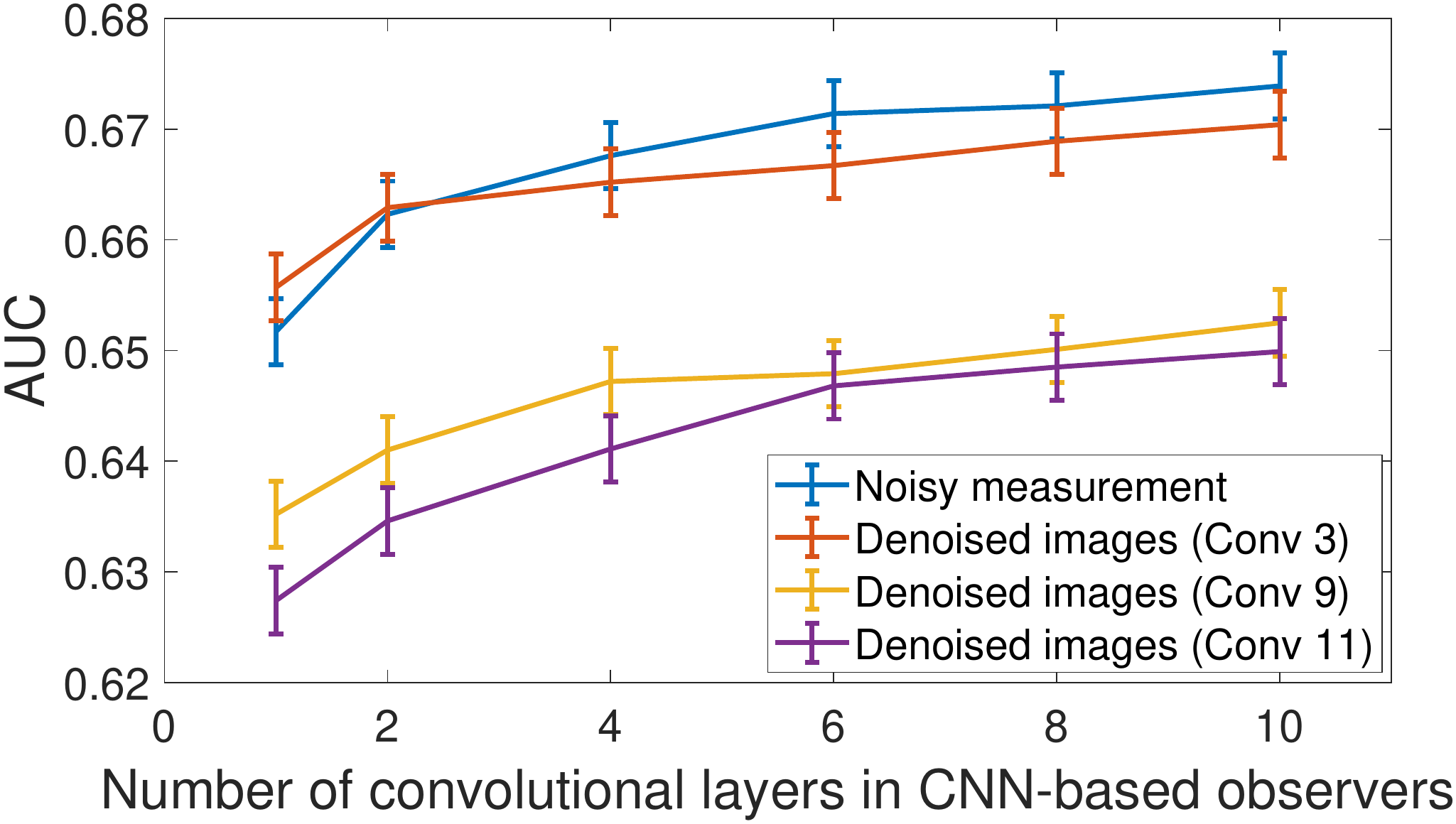}
	\includegraphics[width=0.46\textwidth]{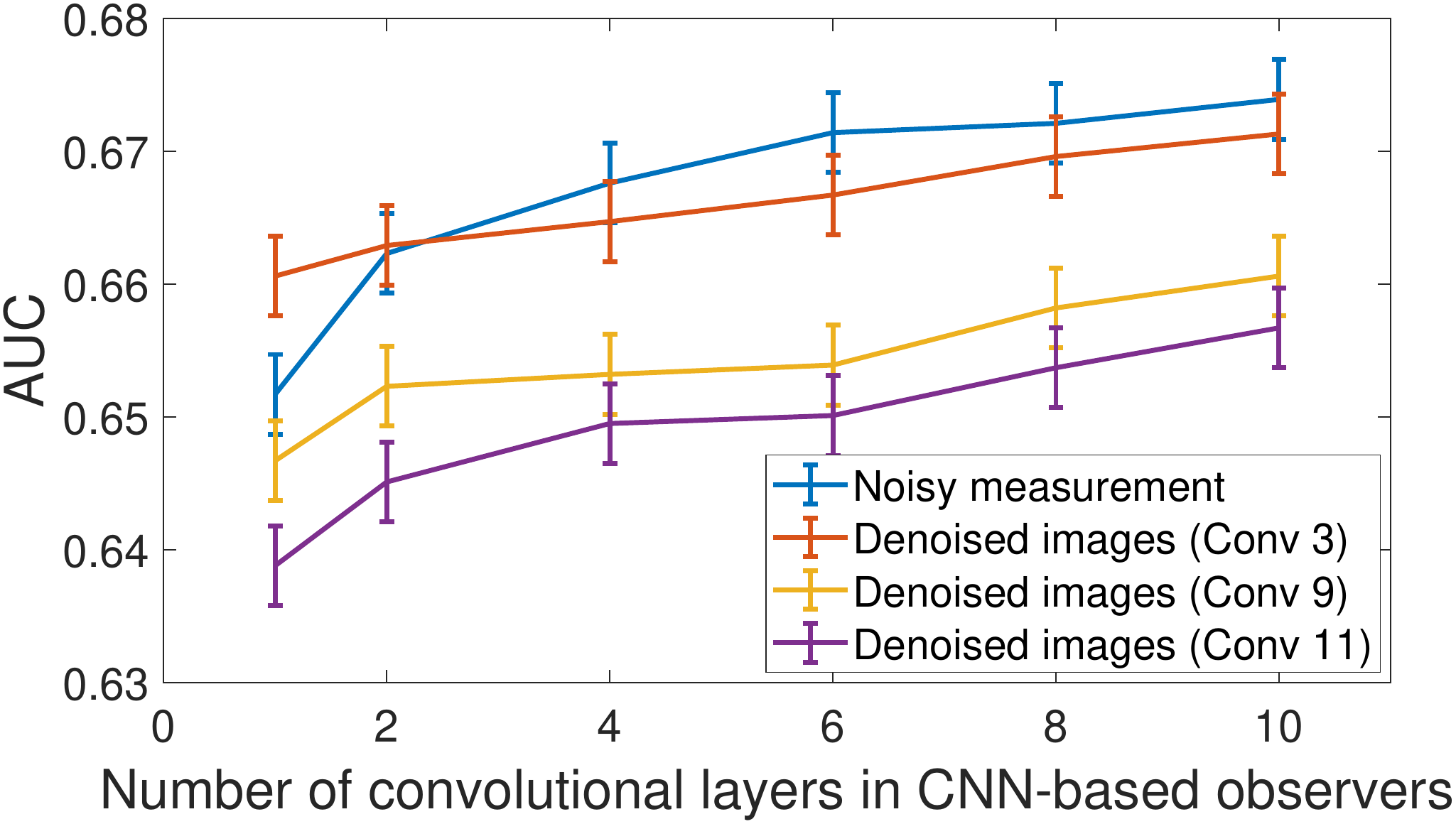}
	\caption{The performance of the CNN-based observers with different number of convolutional layers  acting on the original noisy image and the outputs of two non-linear denoising networks were compared.
	The upper panel shows the results on the CNN-based nonlinear denoising networks; 
	The lower panel shows the results on the ResNet-based nonlinear denoising networks.
	Note that the y-axis range is clipped for display purposes.}
	\label{fig:results2}
\end{figure}

The results shown in Fig.~\ref{fig:results2}
reveal, as expected, that  the performance of the CNN-based observer increases with observer network depth.
More interestingly,
the detection performance of the shallow CNN-based observer with 3 layers on the original noisy images was worse than that on the images denoised by a non-linear denoising network that also had 3 convolutional layers. 
This represented a situation in which the denoising operation resulted in 
improved signal detection performance. 

As observed and discussed above in Section\ \ref{sec:result_a}, the use of deeper denoising networks
resulted in a stronger degradation in signal detection performance  as compared
to use of shallower networks for the NOs considered. 
Additionally, according to data processing inequality~\cite{beaudry2012intuitive}, 
it is known that the performance of an IO cannot be increased via image processing operations such as denoising.
As such, it is to be expected that the performance of the CNN-IO on the original noisy image data
will not be improved by use of any denoising operation.   These factors suggest that the
extent to which a denoising operation will improve signal detection performance depends, in a complicated way on (at least) the following: 1) the extent to which the denoising operation degrades the image statistics that are employed by a given NO for a specified inference; and 2) the extent to which the NO approximates the IO.

\section{Summary and Discussion}
\label{sec:discussion}

In this work, the performance of DNN-based denoising methods was evaluated 
by use of task-based IQ measures.
Specifically,  binary  signal  detection  tasks  under SKE/BKS  conditions were considered. 
The performance of the IO and common linear NOs were quantified 
to assess the impact of the denoising operation on task performance. 
This study was motivated by the scarcity of works that have evaluated such modern denoising methods by use of objective methods.

The numerical results showed that, in the cases considered, 
the denoising operation can result in a loss of task-relevant information.
 Moreover, it was observed that while increasing the depth of the denoising network improved RMSE and SSIM, it
 resulted in a decrease in NO performance.  This is consistent with the well-known fact that
 physical IQ measures may not always correlate with task-based ones \cite{myers1985effect}. This result also suggests  
 that  the mantra ``deep is better" should be qualified and may not always hold true for objective IQ measures.
 The considered networks were analyzed to gain insights into the observed behavior and it was found that the denoising operation resulted in ill-conditioned covariance matrices.  As such,   denoising networks, while 
 seeking to minimize  a traditional (non-task-based) loss function, have the potential to degrade the image statistics that are important for signal detection.

Conditions under which the considered denoising operations could improve NO performance were also investigated.
In the presented studies, it was observed that a shallow denoising network could improve the performance of
a shallow CNN-based observer. When the depth of either the denoising or observer networks increased, the benefit 
of denoising was lost and NO performance was degraded.
This suggests that the impact of denoising on signal detection performance depends, in a complicated way, on the specification of the
denoising network, task, and the NO. As such, there is an urgent need to objectively evaluate new DNN-based denoising methods.

There remain numerous important topics for future investigation. 
The binary SKE  detection task   considered in this study is simplistic relative to many real-word clinical tasks. 
It will be important to consider more complicated tasks that involve signal variability
and hybrid tasks that involve detection and estimation \cite{clarkson2007estimation}.
The study design presented can also readily be applied to assess alternative DNN-based denoising methods that use varying network architectures and loss functions. 
Ultimately, it will be critical to conduct human reader studies to assess the utility
of new DNN-based denoising methods for specific clinical tasks.
 
Finally, the presented results will motivate the development of new approaches to establishing
DNN-based denoising methods that mitigate the loss of task-relevant information by incorporating task-relevant information
in the training strategy.


\bibliography{denoising} 
\bibliographystyle{IEEETran}

\end{document}